\documentclass[twoside,twocolumn,10pt]{article}
\usepackage[pdftex]{color}
\usepackage[bottom,marginal,stable,multiple]{footmisc}
\usepackage{wscg}
\usepackage{mathtools}
\usepackage[pdftex]{graphicx}
\graphicspath{{figures/}{plots/}}
\usepackage{subfigure}
\usepackage{caption}
\DeclareCaptionType{copyrightbox}
\usepackage{nopageno}
\usepackage[numbers]{natbib}
\usepackage{natbibspacing}
\usepackage{tabu}
\usepackage{booktabs}
\usepackage{flushend}
\usepackage[pdfusetitle]{hyperref}

\makeatletter
\def\affiliation#1{\def\@affiliation{#1}}
\def\@maketitle{%
  \begin{center}%
  \let \footnote \thanks
    \sffamily
    {\fontsize{16pt}{19.2pt} \bfseries \@title \par}%
    \vskip 1.0em%
    {%
      \lineskip .5em%
      \begin{tabular}[t]{c}%
        \@author
      \end{tabular}%
      \vskip 0.5em%
      \@affiliation%
      \par}%
  \end{center}%
  \par
  \vskip 0.5em}
\makeatother

\title{Accelerating Discrete Wavelet Transforms\\on Parallel Architectures}
\author{David Barina \and Michal Kula \and Michal Matysek \and Pavel Zemcik}
\affiliation{Centre of Excellence IT4Innovations\\Faculty of Information Technology\\Brno University of Technology\\Bozetechova 1/2, Brno\\Czech Republic\\\{ibarina,ikula,imatysek,zemcik\}@fit.vutbr.cz}

\newcommand*\Figure[1]{Figure~\ref{#1}}
\newcommand*\Table[1]{Table~\ref{#1}}

\let\P\undefined
\newcommand{\U}[1][]{{\mathrm{U}_{#1}}}
\newcommand{\P}[1][]{{\mathrm{P}_{#1}}}
\newcommand{\V}[1][]{{\mathrm{V}_{#1}}}

\newcommand\even[1]{{#1}^{\!(e)}}
\newcommand\odd[1]{{#1}^{\!(o)}}

\def\ww{2.8em}
\newcommand\w[1]{\makebox[\ww]{$#1$}}

\newcommand\T[1][]{\mathrm{T}_{#1}}
\let\S\undefined
\newcommand\S[1][]{\mathrm{S}_{#1}}

\newcommand\N[1][]{\mathrm{N}_{#1}}
\newcommand\NN[1][]{\mathrm{\mathbf{N}}_{#1}}

\begin{document}

\twocolumn[{\csname @twocolumnfalse\endcsname

\maketitle

\begin{abstract}
The \mbox{2-D} discrete wavelet transform (DWT) can be found in the heart of many image-processing algorithms.
Until recently, several studies have compared the performance of such transform on various shared-memory parallel architectures, especially on graphics processing units (GPUs).
All these studies, however, considered only separable calculation schemes.
We show that corresponding separable parts can be merged into non-separable units, which halves the number of steps.
In addition, we introduce an optional optimization approach leading to a reduction in the number of arithmetic operations.
The discussed schemes were adapted on the OpenCL framework and pixel shaders, and then evaluated using GPUs of two biggest vendors.
We demonstrate the performance of the proposed non-separable methods by comparison with existing separable schemes.
The non-separable schemes outperform their separable counterparts on numerous setups, especially considering the pixel shaders.
\end{abstract}

\subsection*{Keywords}
Discrete wavelet transform, Image processing, Synchronization, Graphics processors

\vspace*{1.0\baselineskip}
}]

\section{Introduction}
\label{sec:Introduction}

\copyrightspace

The discrete wavelet transform became a very popular image processing tool in last decades.
A widespread use of this transform has resulted in a development of fast algorithms on all sorts of computer systems, including shared-memory parallel architectures.
At present, the GPU is considered as a typical representative of such parallel architectures.
In this regard, several studies have compared the performance of various \mbox{2-D} DWT computational approaches on GPUs.
All of these studies are based on separable schemes, whose operations are oriented either horizontally or vertically.
These schemes comprise the convolution and lifting.
The lifting requires fewer arithmetic operations as compared with the convolution, at the cost of introducing some data dependencies.
The number of operations should be proportional to a transform performance.
However, also the data dependencies may form a bottleneck, especially on shared-memory parallel architectures.

In this paper, we show that the fastest scheme for a given architecture can be obtained by fusing the corresponding parts of the separable schemes into new structures.
Several new non-separable schemes are obtained in this way.
More precisely, the underlying operations of these schemes can be associated with neither horizontal nor vertical axes.
In addition, we present an approach where each scheme can be adapted to a particular platform in order to reduce the number of operations.
This possibility was completely omitted in existing studies.
Our reasoning is supported by extensive experiments on GPUs using OpenCL and pixel shaders (fragment shaders in OpenGL terminology).
The presented schemes are general, and they are not limited to any specific type of DWT.
To clarify the situation, they all compute the same values.

The rest of this paper is organized as follows.
Section~\nameref{sec:background} formally introduces the problem definition.
Section~\nameref{sec:related-work} briefly presents the existing separable approaches.
Section~\nameref{sec:proposed-schemes} presents the proposed non-separable schemes.
Section~\nameref{sec:improvements} discusses the optimization approach that reduces the number of operations.
Section~\nameref{sec:performance} evaluates the performance on GPUs in the pixel shaders and OpenCL framework.
Eventually, Section~\nameref{sec:conclusion} closes the paper.
This section is followed by Section~\nameref{sec:appendix} for readers not familiar with signal-processing notations.

\section{Background}
\label{sec:background}

Since the separable schemes are built on the one-dimensional transform, a widely-used $z$-transform is used for the description of underlying FIR filters.
The transfer function of the filter $\left( g_k \right)$ is the polynomial
\begin{align*}
	G(z) = \sum_{k} \, g_k \, z^{-k} \text{,}
\end{align*}
where the $k$ refers to the time axis.
Below in the text, the one-dimensional transforms are used in conjunction with two-dimensional signals.
For this case, the transfer function of the filter $\left( g_{k_m,k_n} \right)$ is defined as the bivariate polynomial
\begin{align*}
	G(z_m,z_n) = \sum_{k_m} \sum_{k_n} \, g_{k_m,k_n} \, z_m^{-k_m} z_n^{-k_n} \text{,}
\end{align*}
where the subscript $m$ refers to the horizontal axis and $n$ to the vertical one.
The $ G^*(z_m,z_n) = G(z_n,z_m) $ is a polynomial transposed to a polynomial $ G(z_m,z_n) $.
A shortened notation G is only written in order to keep the notation readable.

A discrete wavelet transform is a signal-processing tool which is suitable for the decomposition of a signal into low-pass and high-pass components.
In detail, the single-scale transform splits the input signal into two components, according to a parity of its samples.
Therefore, the DWT is described by $2 \times 2$ matrices.
As shown by Mallat \cite{Mallat1989}, the transform can be computed by a pair of filters followed by subsampling by a factor of 2.
The filters are referred to as $\mathrm{G}_0, \mathrm{G}_1$.
The transform can also be represented by the polyphase matrix
\begin{align}
	\label{eqn:convolution}
	\begin{bmatrix}
		\w{\odd{\mathrm{G}_1}} & \even{\mathrm{G}_1} \\
		\odd{\mathrm{G}_0} & \w{\even{\mathrm{G}_0}}
	\end{bmatrix}
	\text{,}
\end{align}
where the polynomials $\even{\mathrm{G}}$ and $\odd{\mathrm{G}}$ refer to the even and odd terms of $\mathrm{G}$.
This polyphase matrix defines the convolution scheme.
To avoid misunderstandings, it is necessary to say that, in this paper, column vectors are transformed to become another columns.
For example, $\mathrm{y} = \mathrm{M} \mathrm{x}$ and $\mathrm{y} = \mathrm{M}_2 \mathrm{M}_1 \mathrm{x} $ are transforms represented by one and two matrices, respectively.
Following the algorithm by Sweldens \cite{Sweldens1996,Daubechies1998}, the convolution scheme in (\ref{eqn:convolution}) can be factored into a sequence
\begin{align}
	\label{eqn:lifting}
	\prod_{k}
	\begin{bmatrix}
		1 & \U{}^{(k)} \\
		0 & 1
	\end{bmatrix}
	\begin{bmatrix}
		1 & 0 \\
		\P{}^{(k)} & 1
	\end{bmatrix}
\end{align}
of $K$ pairs of short filterings, known as the lifting scheme.
The filters employed in (\ref{eqn:lifting}) are referred to as the lifting steps.
Usually, the first step $\P{}^{(k)}$ in the $k$th pair is referred to as the predict and the second one $\U{}^{(k)}$ as the update.
The lifting scheme reduces the number of operations by up to half.
Since this paper is mostly focused on a single pair of steps, the superscript $(k)$ is omitted in the text below.
Note that the number of operations is calculated as the number of distinct (in a column) terms of all polynomials in all matrices, excluding units on diagonals.

Considering the shared-memory parallel architectures, the processing of single or several samples is mapped to independent processing units.
In order to avoid race conditions during data exchange, the units must use some synchronization method (barrier).
In the lifting scheme, the barriers are required before the lifting steps.
In the convolution scheme, the barrier is only required before starting the calculation.
In this paper, the barriers are indicated by the $|$ symbol.
For example, $\mathrm{M}_2 | \mathrm{M}_1 $ are two adjacent lifting steps separated by the barrier.
For simplicity, the number of barriers is also called the number of steps in the text below.

\begin{figure}[h]
	\hspace*{\fill}%
	\subfigure{\includegraphics{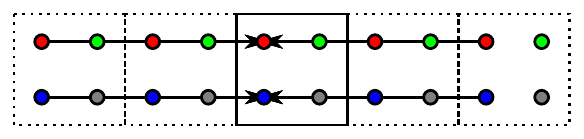}}%
	\hspace*{\fill}%
	\\%
	\hspace*{\fill}%
	\subfigure{\includegraphics{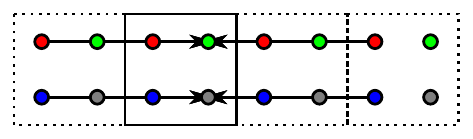}}%
	\hspace*{\fill}%
	\caption{
		Horizontal part of the separable convolution scheme for the CDF\,9/7 wavelet.
		Two appropriately chosen pairs of matrix rows are depicted in separate subfigures.
		The arrows are pointing to the destination operand and denote a multiply--accumulate operation, with multiplication by a real constant.
		The arrows in the same row overlap.
	}
	\label{fig:dataflow-Separable-Convolution}
\end{figure}

The \mbox{2-D} transform is defined as a tensor product of \mbox{1-D} transforms.
Consequently, the transform splits the signal into a quadruple of wavelet coefficients.
Therefore, the \mbox{2-D} DWT is described by $4 \times 4$ matrices.
See Section~\nameref{sec:appendix} for details.
Following the pioneering paper of Mallat \cite{Mallat1989}, the \mbox{1-D} transforms are applied in both directions sequentially.
By its nature, this scheme can be referred to as the separable convolution.
The calculations in a single direction are performed in a single step.
This means two steps for the two dimensions.
The scheme can formally be described as
\begin{align*}
	\NN[]^V \, \big| \, \NN[]^H \, \big| \text{,}
\end{align*}
where $\NN^H$ and $\NN^V$ are \mbox{1-D} transforms in horizontal and in vertical direction.
For the well-known Cohen-Daubechies-Feauveau (CDF) wavelet with 9/7 samples, such as used in the JPEG 2000 standard, these matrices are graphically illustrated in \Figure{fig:dataflow-Separable-Convolution}.
Here, only the horizontal part is shown.
Particularly, the filters in the figure are of sizes 9 and 7 taps.
The \parbox{\wd0}{\hbox{\includegraphics{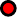}}}, \parbox{\wd0}{\hbox{\includegraphics{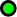}}}, \parbox{\wd0}{\hbox{\includegraphics{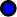}}}, and \parbox{\wd0}{\hbox{\includegraphics{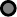}}} circles represent the quadruple of wavelet coefficients.
Figures shown are for illustration purpose only.

\pagebreak

\begin{figure}[h]
	\hspace*{\fill}%
	\subfigure[${\T[\P]^H}$]{\includegraphics{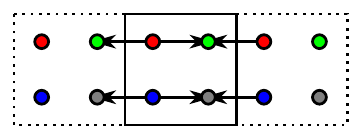}}%
	\hspace*{\fill}
	\subfigure[${\S[\U]^H}$]{\includegraphics{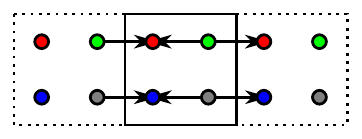}}%
	\hspace*{\fill}%
	\caption{
		The horizontal part of the separable lifting scheme for the CDF wavelets.
	}
	\label{fig:dataflow-Separable-Lifting}
\end{figure}

Another scheme used for \mbox{2-D} transform is the separable lifting.
Similarly to the previous case, the predict and update lifting steps can be applied in both directions sequentially.
Moreover, horizontal and vertical steps can be arbitrarily interleaved thanks to the linear nature of the filters.
Therefore, the scheme is defined as
\begin{align*}
	\S[\U]^V \, \big| \, \S[\U]^H \, \big| \, \T[\P]^V \, \big| \, \T[\P]^H \, \big| \text{,}
\end{align*}
wherein the predict steps $\T$ always precede the update steps $\S$.
The above mapping corresponds to a single $\P$ and $\U$ pair of lifting steps.
For multiple pairs, the scheme is separately applied to each such pair.
In order to describe \mbox{2-D} matrices, the lifting steps must be extended into two dimensions as
\begin{align*}
	\begin{bmatrix}
		\+G^{\phantom{*}} \\
		\+G^*
	\end{bmatrix} = \begin{bmatrix}
		G^{\phantom{*}}(z_m,z_n) \\
		G^*(z_m,z_n)
	\end{bmatrix} = \begin{bmatrix}
		G(z_m) \\
		G(z_n)
	\end{bmatrix} \text{.}
\end{align*}
Then, the individual steps are defined as
\begin{align*}
	{\T[\P]^H} & =
	\begin{bmatrix}
		\w{1} & 0     & 0     & 0     \\
		\P    & \w{1} & 0     & 0     \\
		0     & 0     & \w{1} & 0     \\
		0     & 0     & \P    & \w{1} \\
	\end{bmatrix} \text{,}
	\\
	{\T[\P]^V} & =
	\begin{bmatrix}
		\w{1} & 0     & 0     & 0     \\
		0     & \w{1} & 0     & 0     \\
		\P^*  & 0     & \w{1} & 0     \\
		0     & \P^*  & 0     & \w{1} \\
	\end{bmatrix} \text{,}
	\\
	{\S[\U]^H} & =
	\begin{bmatrix}
		\w{1} & \U    & 0     & 0     \\
		0     & \w{1} & 0     & 0     \\
		0     & 0     & \w{1} & \U    \\
		0     & 0     & 0     & \w{1} \\
	\end{bmatrix} \text{,}
	\\
	{\S[\U]^V} & =
	\begin{bmatrix}
		\w{1} & 0     & \U^*  & 0     \\
		0     & \w{1} & 0     & \U^*  \\
		0     & 0     & \w{1} & 0     \\
		0     & 0     & 0     & \w{1} \\
	\end{bmatrix} \text{.}
\end{align*}
For the CDF wavelets, the matrices are also illustrated in \Figure{fig:dataflow-Separable-Lifting}, again showing the horizontal part only.

\bigskip

\newpage

\section{Related Work}
\label{sec:related-work}

This section briefly reviews papers that motivated our research.
So far, several papers have compared the performance of the separable lifting and separable convolution schemes on GPUs.
Especially, Tenllado \textit{et al.} \cite{Tenllado2008} compared these schemes on GPUs using pixel shaders.
The authors mapped data to \mbox{2-D} textures, constituted by four floating-point elements.
They concluded that the separable convolution is more efficient than the separable lifting scheme in most cases.
They further noted that fusing several consecutive kernels might significantly speed up the execution, even if the complexity of the resulting fused pixel program is higher.

Kucis \textit{et al.} \cite{Kucis2014} compared the performance of several recently published schedules for computing the \mbox{2-D} DWT using the OpenCL framework.
All of these schedules use separable schemes, either the convolution or lifting.
In more detail, the work compares a convolution-based algorithm proposed in \cite{Galiano2011} against several lifting-based methods \cite{Blazewicz2012,Laan2011} in the horizontal part of the transform.
The authors concluded that the lifting-based algorithm of Blazewicz \textit{et al.} \cite{Blazewicz2012} is the fastest method.
Furthermore, Laan \textit{et al.} \cite{Laan2011} compared the performance of their separable lifting-based method against a convolution-based method.
They concluded that the lifting is the fastest method.
The authors also compared the performance of implementations in CUDA and pixel shaders, based on the work of Tenllado \cite{Tenllado2008}.
The CUDA implementation proved to be the faster choice.
In this regard, the authors noted that a speedup in CUDA occurs because the CUDA effectively makes use of on-chip memory.
This use is not possible in pixel shaders, which exchange the data using off-chip memory.
Other important separable approaches can be found in \cite{Matela2009,Galiano2013,Song2014,Quan2016}.

This paper is based on the previous works in \cite{Barina2016,Kula2016}.
In those works, we introduced several non-separable schemes for calculation of \mbox{2-D} DWT.
However, we have not considered important structures, such as polyconvolutions.
We contribute this consideration with this paper.
Moreover, differences and similarities between the separable schemes and their non-separable counterparts are homogeneously discussed here.
All these schemes are also thoroughly analyzed and evaluated.

Considering the present papers, we see that a possible fusion of separable parts into new non-separable structures is not considered.
Therefore, we investigate on this promising technique in the following sections.

\newpage

\twocolumn[{\csname @twocolumnfalse\endcsname
\begin{minipage}[b]{\textwidth}
	\hspace*{\fill}%
	{\includegraphics{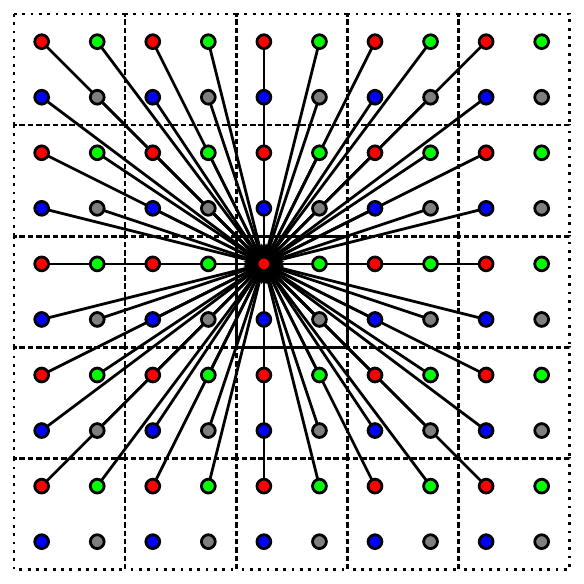}}%
	\hspace*{\fill}%
	{\includegraphics{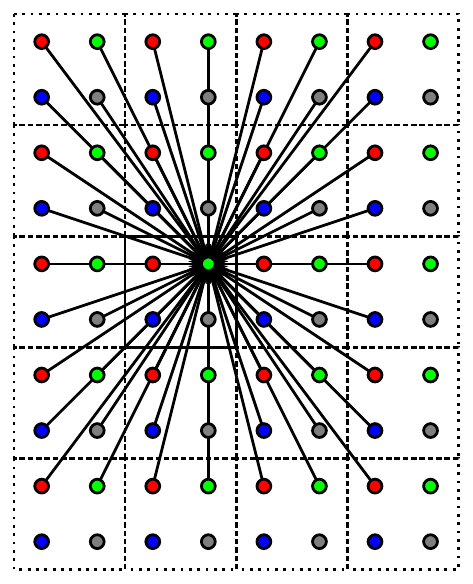}}%
	\hspace*{\fill}%
	\null\\%
	\hspace*{\fill}%
	{\includegraphics{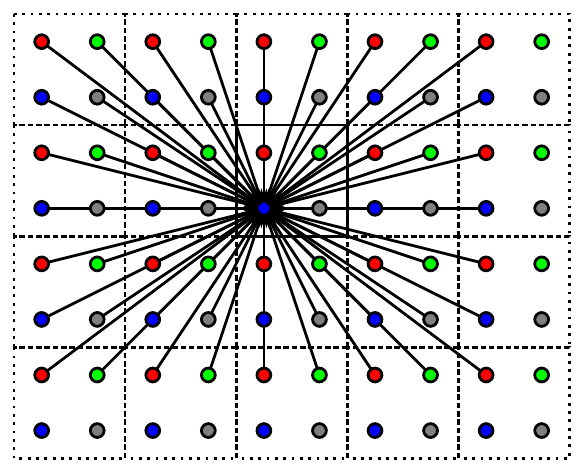}}%
	\hspace*{\fill}%
	{\includegraphics{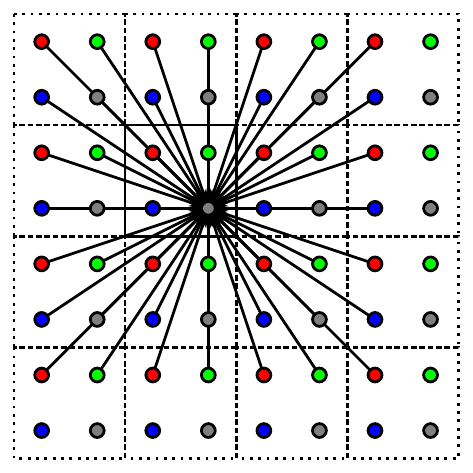}}%
	\hspace*{\fill}%
	\null%
	\captionof{figure}{
		Non-separable convolution scheme for the CDF\,9/7 wavelet.
		The individual rows of $\N$ are depicted in separate subfigures.
		The sizes are from top to bottom and left to right: $9\times9$, $7\times9$, $9\times7$, $7\times7$.
	}
	\label{fig:dataflow-Non-Separable-Convolution}
\end{minipage}
\vspace*{1.0\baselineskip}
}]

\section{Proposed Schemes}
\label{sec:proposed-schemes}

\smallskip

As stated above, the existing approaches did not study the possibility of a partial fusion of lifting polyphase matrices.
This section presents three alternative non-separable schemes for the calculation of the \mbox{2-D} transform.
The contribution of this paper starts with this section.
To avoid confusion, please note that the proposed schemes compute the same values as the original ones.

\bigskip

The non-separable convolution scheme is a counterpart to the separable convolution.
Unlike the separable scheme, all horizontal and vertical calculations are performed in a single step
\begin{align*}
	\NN \, \big|\text{,}
\end{align*}
where $\NN = \NN^V \NN^H$ is a product of \mbox{1-D} transforms in horizonal and vertical directions.
The drawback of this scheme is that it requires the highest number of arithmetic operations.
For the CDF\,9/7 wavelet, the matrix is graphically illustrated in \Figure{fig:dataflow-Non-Separable-Convolution}.
Here, the \mbox{2-D} filters are of sizes $9\times9$, $7\times9$, $9\times7$, and $7\times7$.
These sizes make the calculation computationally demanding.
Aside from the GPUs, this approach was earlier discussed in Hsia \textit{et al.} \cite{Hsia2009}.

\newpage

\begin{figure}[h]
	\hspace*{\fill}%
	\subfigure{\includegraphics{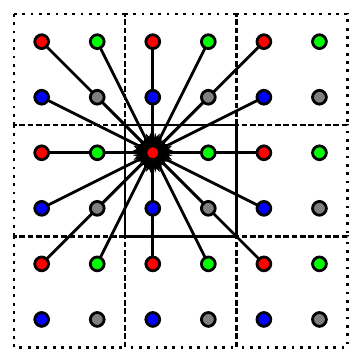}}%
	\hspace*{\fill}%
	\subfigure{\includegraphics{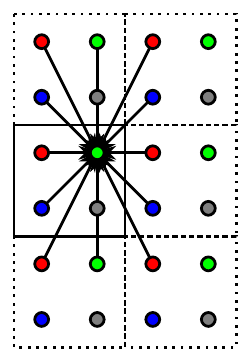}}%
	\hspace*{\fill}%
	\\
	\hspace*{\fill}%
	\subfigure{\includegraphics{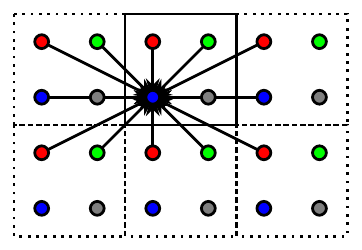}}%
	\hspace*{\fill}%
	\subfigure{\includegraphics{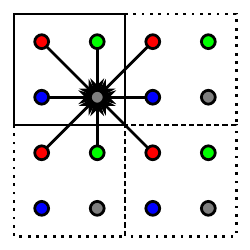}}%
	\hspace*{\fill}%
	\null%
	\caption{
		Non-separable polyconvolution scheme for the CDF wavelets.
		The individual rows of $\N$ are depicted in separate subfigures.
	}
	\label{fig:dataflow-Non-Separable-Polyconvolution}
\end{figure}

In order to reduce computational complexity, it would be a good idea to construct some smaller non-separable steps.
Indeed, the non-separable convolution can be broken into smaller units, referred here to as the non-separable polyconvolutions.
For a single pair of lifting steps, the scheme follows from the mapping
\begin{align*}
	\N[\P,\U] \, \big| \text{,}
\end{align*}
where
\begin{align*}
	{\N[\P,\U]} =
	\begin{bmatrix}
		\w{\V^*\V} & \V^*\U   & \U^*\V & \U^*\U \\
		\V^*\P     & \w{\V^*} & \U^*\P & \U^*   \\
		\P^*\V     & \P^*\U   & \w{\V} & \U     \\
		\P^*\P     & \P^*     & \P     & \w{1}  \\
	\end{bmatrix}
\end{align*}
and $\V = \P\U + 1$.
For the CDF wavelets, the scheme is graphically illustrated in \Figure{fig:dataflow-Non-Separable-Polyconvolution}.
In this case, the employed filters are of sizes $5\times5$, $3\times5$, $5\times3$, and $3\times3$.
Note that only half of the operations are required specifically for the CDF\,9/7 wavelet, compared to the {non-separable convolution}.
For the sake of completeness, it should be pointed out that it is also possible to formulate the separable polyconvolution scheme.
In our experiments, this one was however not proven to be useful concerning the performance.

\pagebreak

\begin{figure}[h]
	\hspace*{\fill}%
	\subfigure[${\T[\P]}$]{\includegraphics{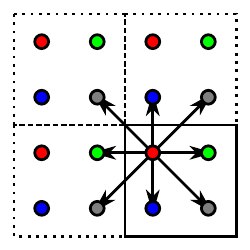}}%
	\hspace*{\fill}%
	\subfigure[${\T[\P]}$]{\includegraphics{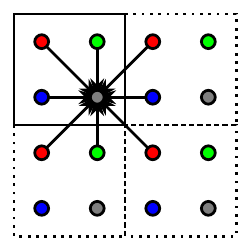}}%
	\hspace*{\fill}%
	\\
	\hspace*{\fill}%
	\subfigure[${\S[\U]}$]{\includegraphics{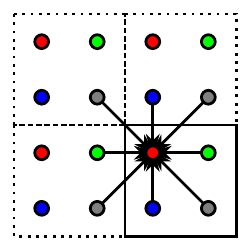}}%
	\hspace*{\fill}%
	\subfigure[${\S[\U]}$]{\includegraphics{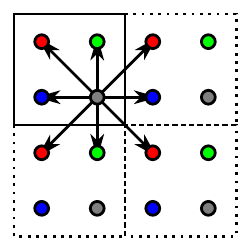}}%
	\hspace*{\fill}%
	\null%
	\caption{
		{Non-separable lifting} scheme for the CDF wavelets.
	}
	\label{fig:dataflow-Non-Separable-Lifting}
\end{figure}

By combining the corresponding horizontal and vertical steps of the separable lifting scheme, the non-separable lifting scheme is formed.
The number of operations has slightly been increased.
The scheme consists of a spatial predict and spatial update step, thus two steps in total for each pair of the original lifting steps.
Formally, for each pair of $\P$ and $\U$, the scheme follows from
\begin{align*}
	\S[\U] \, \big| \, \T[\P] \, \big| \text{,}
\end{align*}
where
\begin{align*}
	{\T[\P]} & =
	\begin{bmatrix}
		\w{1}  & 0     & 0     & 0     \\
		\P     & \w{1} & 0     & 0     \\
		\P^*   & 0     & \w{1} & 0     \\
		\P\P^* & \P^*  & \P    & \w{1} \\
	\end{bmatrix} \text{,}
	\\
	{\S[\U]} & =
	\begin{bmatrix}
		\w{1} & \U    & \U^*  & \U\U^* \\
		0     & \w{1} & 0     & \U^*   \\
		0     & 0     & \w{1} & \U     \\
		0     & 0     &     0 & \w{1}  \\
	\end{bmatrix} \text{.}
\end{align*}
Note that the spatial filters in $\P\P^*$ and $\U\U^*$ may be computationally demanding, depending on their sizes.
However, the situation is always better than in the previous two cases.
For the CDF wavelets, the scheme is graphically illustrated in \Figure{fig:dataflow-Non-Separable-Lifting}.

\section{Optimization Approach}
\label{sec:improvements}

This section presents an optimization approach that reduces the number of operations, while the number of steps remains unaffected.
Such an approach was not covered in existing studies.

\begin{figure}[h]
	\hspace*{\fill}%
	\subfigure[${\T[{\P[0]}]^H}$]{\includegraphics{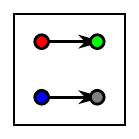}}%
	\hspace*{\fill}%
	\subfigure[${\T[{\P[0]}]^V}$]{\includegraphics{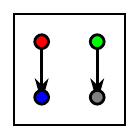}}%
	\hspace*{\fill}%
	\subfigure[${\S[{\U[0]}]^H}$]{\includegraphics{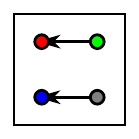}}%
	\hspace*{\fill}%
	\subfigure[${\S[{\U[0]}]^V}$]{\includegraphics{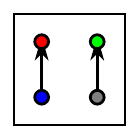}}%
	\hspace*{\fill}%
	\null%
	\caption{
		Separable lifting scheme with the polynomials $\P[0]$ and $\U[0]$.
	}
	\label{fig:the-trick}
\end{figure}

Regardless of the underlying platform, an important observation can be made.
A very special form of the operations guarantees that the processing units never access the results belonging to their neighbors.
These operations comprise only constants.
Since the convolution is a linear operation, the polynomials can be pulled out of the original matrices, and calculated in a different step.
Formally, the original polynomials are split as $\P = \P[0] + \P[1]$ and $\U = \U[0] + \U[1]$.
The $\P[0]$ and $\U[0]$ are constant.
As a next step, the $\P[0]$ and $\U[0]$ are substituted into the separable lifting scheme.
The separable lifting scheme was chosen because it has the lowest number of operations.
This part is illustrated in \Figure{fig:the-trick}.
In contrast, the $\P[1]$ and $\U[1]$ are kept in original schemes.
These two steps are then computed without any barrier.
The observation is further exploited to adapt schemes for a particular platform.

Now, the improved schemes for the shaders and OpenCL are briefly described.
These schemes exploit the above-described observation with the polynomials $\P[0]$ and $\U[0]$ .
On recent GPUs, OpenCL schemes also omit memory barriers due to the \mbox{SIMD-32} architecture.
Note that the non-separable polyconvolution scheme makes sense only when $K>1$, which is the case of the CDF\,9/7 wavelet.
Implementations in the pixel shaders map input and output data to \mbox{2-D} textures.
There is no possibility to retain some results in registers, and the results are exchanged through textures in off-chip memory.
Considering the OpenCL implementations, a data format is not constrained.
The image is divided into overlapping blocks and on-chip memory shared by all threads in a block is utilized to exchange the results.
Additionally, some results are passed in registers.

This paper explores the performance for three frequently used wavelets, namely, CDF\,5/3, CDF\,9/7 \cite{Cohen1992}, and DD\,13/7 \cite{Sweldens1996}.
Their fundamental properties are listed in \Table{tab:parameters-baseline}: number of steps and arithmetic operations.
Note that the number of operations is commonly proportional to a transform performance.
Additionally, the number of steps correspond to the number of synchronizations on parallel architectures, which also form a performance bottleneck.

\begin{table}[h]
	\caption{%
		The total number of steps and arithmetic operations for the optimized schemes.
	}
	\subtable[CDF\,5/3]{%
		\begin{tabu} to \linewidth {X[1.5r]X[l]X[c]X[c]X[c]}
		\toprule
			\multicolumn{2}{c}{\qquad\qquad{}scheme} & steps & \multicolumn{2}{c}{operations} \\
			            ~ &               ~ &     ~ & \multicolumn{1}{c}{OpenCL} & \multicolumn{1}{c}{shaders} \\
		\midrule
			separable     & convolution     &  2 &  20 &  22 \\
			separable     & lifting         &  4 &  16 &  16 \\
		\midrule
			\small{}non\nobreakdash-separable & convolution     &  1 &  23 &  39 \\
			\small{}non\nobreakdash-separable & lifting         &  2 &  18 &  18 \\
		\bottomrule
		\end{tabu}%
	}\\%
	\subtable[CDF\,9/7]{%
		\begin{tabu} to \linewidth {X[1.5r]X[l]X[c]X[c]X[c]}
		\toprule
			\multicolumn{2}{c}{\qquad\qquad{}scheme} & steps & \multicolumn{2}{c}{operations} \\
			            ~ &               ~ &     ~ & \multicolumn{1}{c}{OpenCL} & \multicolumn{1}{c}{shaders} \\
		\midrule
			separable     & convolution     &  2 &  56 &  58 \\
			separable     & polyconv.       &  4 &  40 &  56 \\ 
			separable     & lifting         &  8 &  32 &  32 \\
		\midrule
			\small{}non\nobreakdash-separable & convolution     &  1 & 152 & 200 \\
			\small{}non\nobreakdash-separable & polyconv. &  2 &  46 &  62 \\
			\small{}non\nobreakdash-separable & lifting         &  4 &  36 &  36 \\
		\bottomrule
		\end{tabu}%
	}\\%
	\subtable[DD\,13/7]{%
		\begin{tabu} to \linewidth {X[1.5r]X[l]X[c]X[c]X[c]}
		\toprule
			\multicolumn{2}{c}{\qquad\qquad{}scheme} & steps & \multicolumn{2}{c}{operations} \\
			            ~ &               ~ &     ~ & \multicolumn{1}{c}{OpenCL} & \multicolumn{1}{c}{shaders} \\
		\midrule
			separable     & convolution     &  2 &  60 &  60 \\
			separable     & lifting         &  4 &  32 &  32 \\
		\midrule
			\small{}non\nobreakdash-separable & convolution     &  1 & 203 & 228 \\
			\small{}non\nobreakdash-separable & lifting         &  2 &  50 &  50 \\
		\bottomrule
		\end{tabu}%
	}\\
	\label{tab:parameters-baseline}
\end{table}

\section{Evaluation}
\label{sec:performance}

The experiments in this paper were performed on GPUs of the two biggest vendors NVIDIA and AMD using the OpenCL and pixel shaders.
In these experiments, only a transform performance was measured, usually in gigabytes per second (GB/s).
The host system does not help in the calculation, i.e. with respect to padding or pre/post-processing.
Results for only two GPUs are shown for the sake of brevity: AMD Radeon HD 6970 and NVIDIA Titan X.
Their technical parameters are summarized in \Table{tab:gpus}.

\begin{table}[h]
	\bigskip%
	\caption{
		Specifications of the evaluated GPUs.
	}%
	\small
	\begin{tabu} to \linewidth {l|X[r]X[r]}
		\toprule
		label            &       AMD 6970 &  NVIDIA Titan~X \\
		\midrule
		model            & Radeon~HD~6970 & Titan X (Pascal) \\
		\midrule
		multiprocessors  &             24 &               28 \\
		total processors &         1\,536 &           3\,584 \\
		processor clock  &       880\,MHz &      1\,417\,MHz \\
		performance      & 2\,703\,GFLOPS &  10\,157\,GFLOPS \\
		\midrule
		memory clock     &    1\,375\,MHz &      2\,500\,MHz \\
		bandwidth        &      176\,GB/s &        480\,GB/s \\
		on-chip memory   &        32\,KiB &          96\,KiB \\
		\bottomrule
	\end{tabu}
	\bigskip%
	\label{tab:gpus}
\end{table}

\newpage

The implementations were created using the DirectX HLSL and OpenCL.
The HLSL implementation is used on the NVIDIA Titan X, whereas the OpenCL implementation on the AMD 6970.
The results of the performance comparison are shown in Figures \ref{fig:plots-53}, \ref{fig:plots-97}, and \ref{fig:plots-137}.
The value on the x-axis is the image resolution in kilo/megapixels (kpel or Mpel).
Except for the convolutions for the DD\,13/7 wavelet, the non-separable schemes always outperform their separable counterparts.
For CDF wavelets, having short lifting filters, the non-separable (poly)convolutions have a better performance than the non-separable lifting scheme.
Unfortunately, for the DD\,13/7 wavelet, which is characterized by a high number of operations in lifting filters, the results are not conclusive.
Considering the implementation in pixel shaders, similar results were also achieved on other GPUs, including NVIDIA unified architectures and AMD GPUs based on Graphics Core Next (GCN) architecture.
Whereas for the OpenCL implementation, the non-separable schemes are only proved to be useful for very long instruction word (VLIW) architectures.

\bigskip

Looking at the experiments with the pixel-shader implementations, some transients can be seen at the beginning of the plots (in lower $2$\,Mpel region).
We concluded that these transients are caused by a suboptimal use of cache system, or alternatively by some overhead made by the graphics API.
It should be interesting to show some measures provided by an OpenCL profiler.
Our profiling revealed that the implementations exhibit only an occupancy 95.24\,\%.
This occupancy is caused by making use of 256 threads in OpenCL work groups and due to maximal number 1344 of threads in multiprocessors (256 times 5 work groups is 1280 out of 1344).

\newpage

\twocolumn[{\csname @twocolumnfalse\endcsname
\begin{minipage}[b]{\textwidth}
	\centering%
	{\includegraphics[width=.5\linewidth]{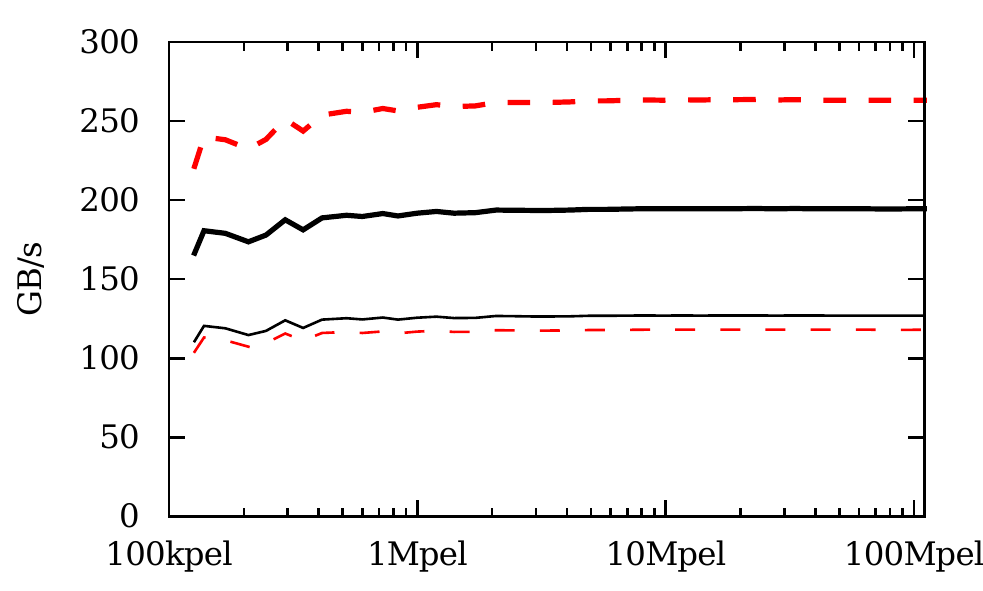}}%
	{\includegraphics[width=.5\linewidth]{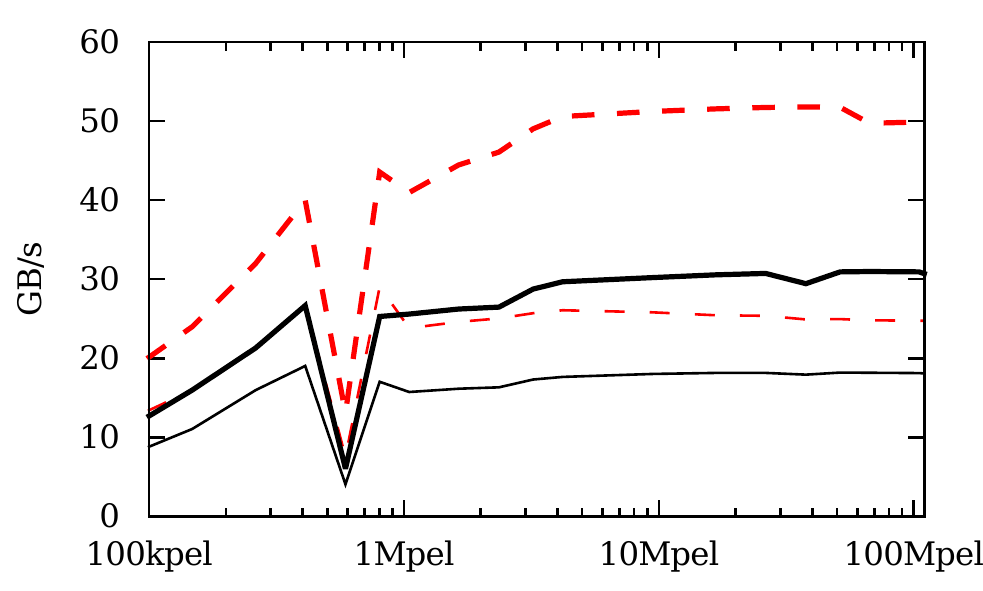}}\\[10pt]%
	\hspace*{\fill}(a) OpenCL\hspace*{\fill}\null\hspace*{\fill}(b) pixel shader\hspace*{\fill}\null\\[10pt]%
	\includegraphics[width=\linewidth]{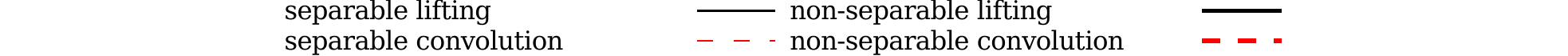}\\[10pt]%
	\captionof{figure}{
		Performance for the CDF\,5/3 wavelet.
	}
	\label{fig:plots-53}
\end{minipage}\\%
\vspace*{1.0\baselineskip}\\%
\begin{minipage}[b]{\textwidth}
	\centering%
	\includegraphics[width=.5\linewidth]{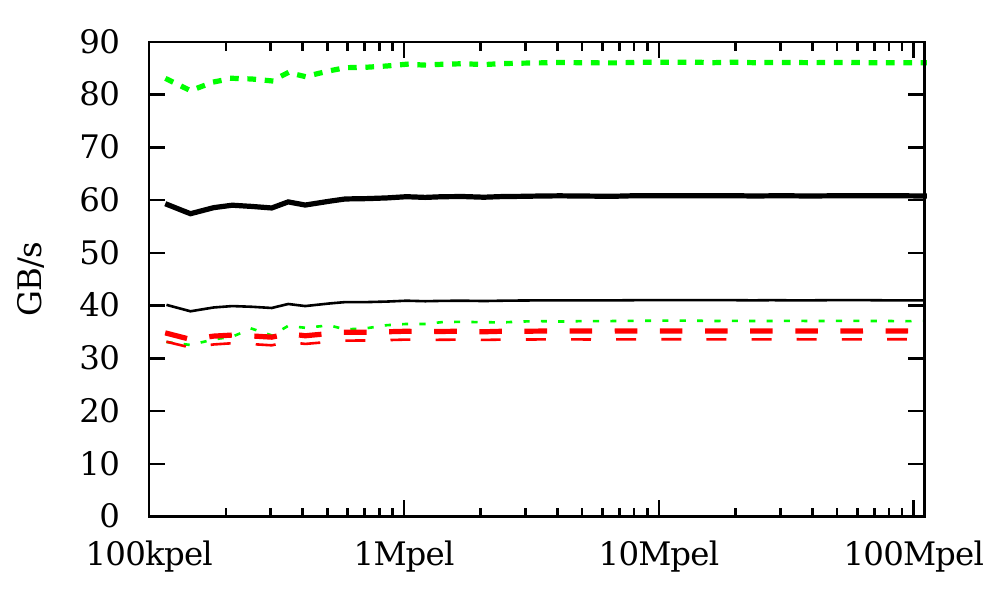}%
	\includegraphics[width=.5\linewidth]{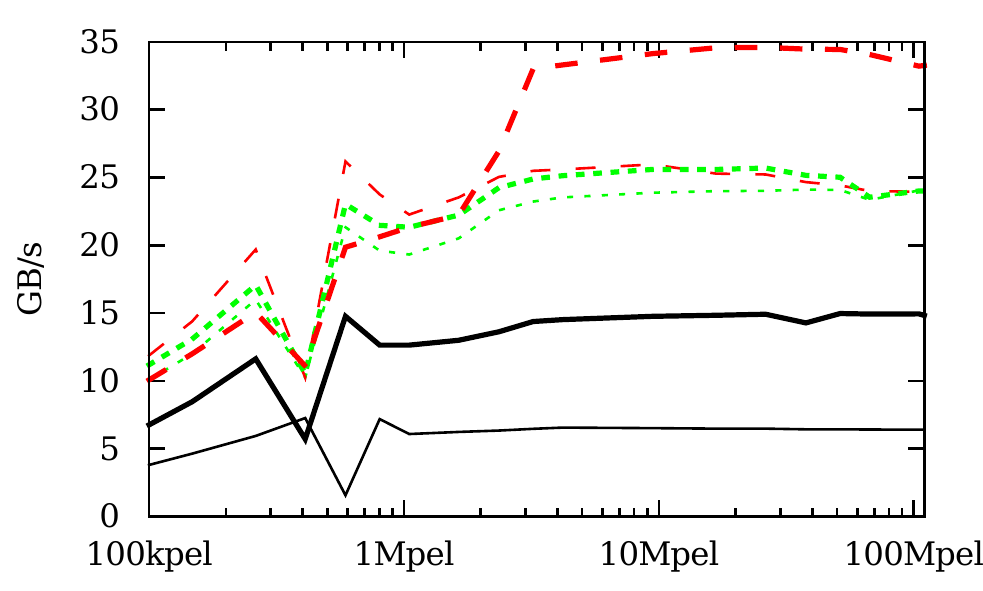}\\[10pt]%
	\hspace*{\fill}(a) OpenCL\hspace*{\fill}\null\hspace*{\fill}(b) pixel shader\hspace*{\fill}\null\\[10pt]%
	\includegraphics[width=\linewidth]{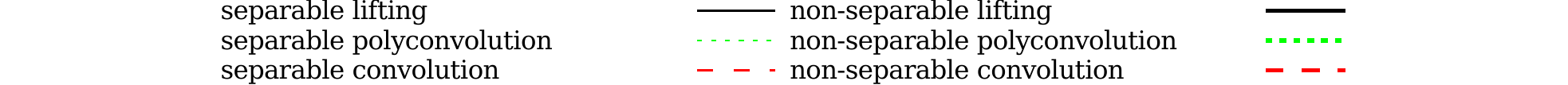}\\[10pt]%
	\captionof{figure}{
		Performance for the CDF\,9/7 wavelet.
	}
	\label{fig:plots-97}
\end{minipage}\\%
\vspace*{1.0\baselineskip}\\%
\begin{minipage}[b]{\textwidth}
	\centering%
	\includegraphics[width=.5\linewidth]{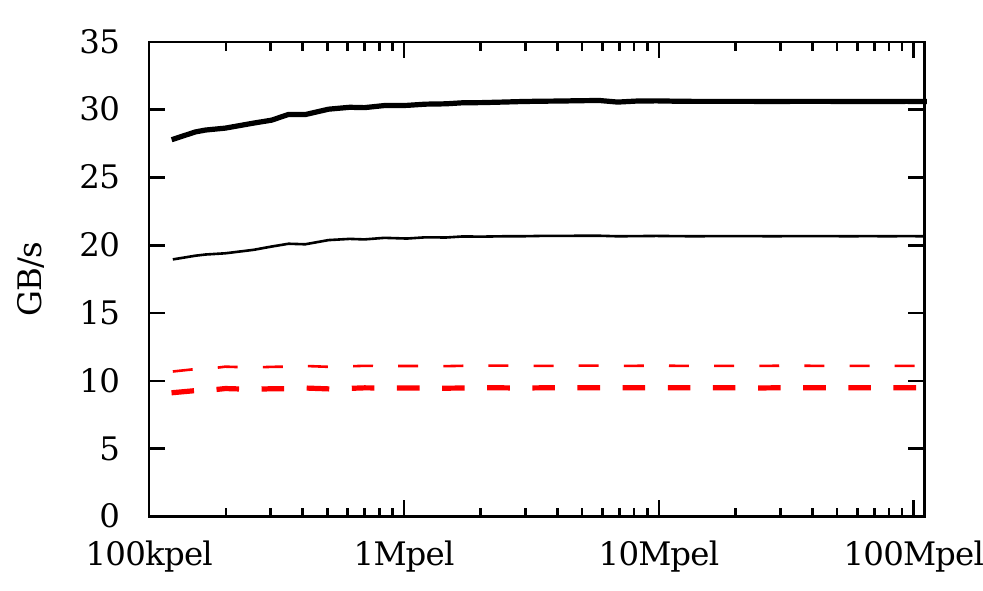}%
	\includegraphics[width=.5\linewidth]{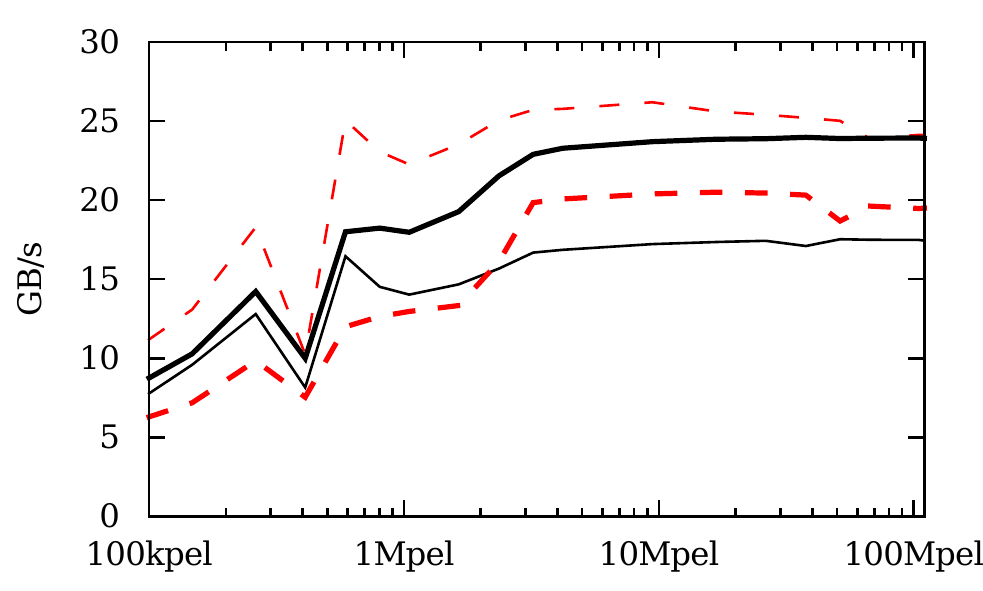}\\[10pt]%
	\hspace*{\fill}(a) OpenCL\hspace*{\fill}\null\hspace*{\fill}(b) pixel shader\hspace*{\fill}\null\\[10pt]%
	\includegraphics[width=\linewidth]{plots/key-53-ink}\\[10pt]%
	\captionof{figure}{
		Performance for the DD\,13/7 wavelet.
	}
	\label{fig:plots-137}
\end{minipage}
\vspace*{1.0\baselineskip}
}]

\section{Conclusions}
\label{sec:conclusion}

This paper presented and discussed several non-separable schemes for the computation of the \mbox{2-D} discrete wavelet transform on parallel architectures, exemplarily on modern GPUs.
As an option, an optimization approach leading to a reduction in the number of operations was presented.
Using this approach, the schemes were adapted on the OpenCL framework and pixel shaders.
The implementations were then evaluated using GPUs of the two biggest vendors.
Considering OpenCL, the schemes exploit features of recent GPUs, such as warping.
For CDF wavelets, the non-separable schemes exhibit a better performance than their separable counterparts on both the OpenCL and pixel shaders.

\medskip

In the evaluation, we reached the following conclusions.
Fusing several consecutive steps of the schemes might significantly speed up the execution, irrespective of their higher complexity.
The non-separable schemes outperform their separable counterparts on numerous setups, especially considering the pixel shaders.
All of the schemes are general and they can be used on any discrete wavelet transform.
In future work, we plan to focus on general-purpose processors and multi-scale transforms.

\paragraph{Acknowledgements}
This work has been supported by
the Ministry of Education, Youth and Sports from the National Programme of Sustainability (NPU II) project IT4Innovations excellence in science (no. LQ1602), and
the Technology Agency of the Czech Republic (TA CR) Competence Centres project V3C -- Visual Computing Competence Center (no. TE01020415).

\newpage

\section*{Appendix}
\label{sec:appendix}

For readers who are not familiar with signal-processing notations, a relationship between polyphase matrices and data-flow diagrams is explained here.
The \mbox{2-D} discrete wavelet transform divides the image into four polyphase components.
Therefore, the $4\times4$ matrices of Laurent polynomials are used to describe the \mbox{2-D} discrete wavelet transform.
These matrices are commonly referred to as the polyphase matrices.
The Laurent polynomials correspond to \mbox{2-D} FIR filters, that define the transform.
In most cases, the transform is described using a sequence of such matrices.
One particular matrix thus defines a step of calculation in this case.

\begin{figure}[h]
	\hspace*{\fill}%
	\subfigure[${\T[\P]^H}$]{\includegraphics{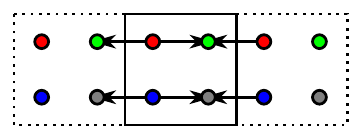}}%
	\hspace*{\fill}
	\caption{
		Visual representation of the polyphase matrix.
		The four polyphase components are represented by color circles.
	}
	\label{fig:dataflow-appendix}
\end{figure}

For example, the matrix
\begin{align*}
	{\T[\P]^H} & =
	\begin{bmatrix}
		\w{1} & 0     & 0     & 0     \\
		\P    & \w{1} & 0     & 0     \\
		0     & 0     & \w{1} & 0     \\
		0     & 0     & \P    & \w{1} \\
	\end{bmatrix}
\end{align*}
maps four polyphase components to another four components, while using two \mbox{2-D} FIR filters represented by the polynomials $\P$.
Moreover, when we substitute a particular polynomial, say $ P(z) = -1/2( 1 + z^{-1} ) $, into the matrix, the mapping gets a specific shape.
Such a substitution illustrated by the data-flow diagram in \Figure{fig:dataflow-appendix}.
The solid arrows correspond to multiplication by $-1/2$ along with subsequent summation.

\newpage

\medskip\medskip
\bibliographystyle{myabbrvnat}
\bibliography{IEEEfull,sources}

\end{document}